\documentclass[journal=jacsat,manuscript=article]{achemso}

\usepackage[version=3]{mhchem} 
\usepackage[T1]{fontenc}       
\usepackage{graphics}
\usepackage[usenames, dvipsnames]{color}
\usepackage{longtable}
\usepackage{multirow}
\usepackage{makecell}
\usepackage{natmove}
\setkeys{acs}{usetitle = true}
\usepackage{amsmath,achemso}
\usepackage{bm}
\setkeys{acs}{maxauthors = 9}

\author{Duy Khanh Nguyen}
\email{nguyenduykhanh@tdtu.edu.vn}
\affiliation{Laboratory of Applied Physics, Advanced Institute of Materials Science, Ton Duc Thang University, Ho Chi Minh City, Vietnam, and Faculty of Applied Science, Ton Duc Thang University, Ho Chi Minh City, Vietnam}
\author{Ngoc Thanh Thuy Tran}
\affiliation{Hierarchical Green-Energy Materials (Hi-GEM) Research Center, National Cheng Kung University, Tainan 701, Taiwan}
\author{Godfrey Gumbs}
\affiliation{Department of Physics and Astronomy, Hunter College of the City University of New York, 695 Park Avenue, New York, New York 10065, USA}
\author{Ming-Fa Lin}
\email{mflin@mail.ncku.edu.tw}
\affiliation {Department of Physics/Hi-GEM/QTC, National Cheng Kung University, Tainan, Taiwan}

\title[An \textsf{achemso} demo]
  {Rich Essential Properties of Si-Doped Graphene}

\abbreviations{DOS, XPS, STM, ARPES, STS}
\keywords{graphene, halogen, first-principles, chemical bonding, energy gap.}

\begin{document}


\begin{abstract}
A theoretical framework, under the first-principles calculations, is developed to fully explore the dramatic changes of essential properties due to silicon-atom chemical modifications on monolayer graphenes. For the Si-chemisorption and Si-substituted graphenes, the guest-atom-diversified geometric structures, the Si- and C-dominated energy bands, the charge transfers, the spatial charge densities, and the van Hove singularities in the atom- and orbital-projected density of states are investigated thoroughly by delicate evaluations and analyses. Such fundamental properties are sufficient to determine the critical physical and chemical pictures, in which the accurate multi-orbital hybridizations are very useful in comprehending the diverse phenomena, e.g., the  C- and Si-co-dominated energy bands, the semiconducting, semimetallic, and metallic behavior, and the existence/absence of Dirac-cone structures. This developing model could be generalized to other emergent layered materials. 
  
\end{abstract}


\section{1. Introduction}

Carbon atoms can form three-dimensional (3D) diamond\cite{1}, 3D graphites\cite{2},  two-dimensional (2D) layered graphenes\cite{3,4}, one-dimensional (1D) graphene nanoribbons\cite{5,6,7}, 1D carbon nanotubes\cite{8,9,10}, zero-dimensional (0D) carbon toroids\cite{11,12}, 0D C$_{60}$-related fullerenes\cite{13}, and 0D carbon onions\cite{14}. The versatile morphologies directly indicate the peculiar chemical bondings, in which all carbon-created systems possess ${sp^2}$-bonding surfaces except for the ${sp^3}$ bondings in diamond. Specifically, the few- and multi-layer graphene systems have been manufactured using the various methods\cite{15,16} since the first experimental observation in 2004 by mechanical exfoliation. Up to now, they clearly exhibit plenty of remarkable fundamental properties due to the hexagonal symmetry, the nanoscaled thickness, and the distinct stacking configurations, such as semiconducting and semi-metallic behaviors\cite{17,18}, anomalous quantum Hall effects\cite{19}, diverse magnetic quantizations\cite{20,21,22,23}, rich Coulomb excitations and decays\cite{24,25,26,27,28}, different magneto-optical selection rules\cite{29,30,31}, the exceedingly high mobility of charge carriers\cite{32,33}, and the largest Young's modulus of materials ever tested\cite{34}.  To induce the novel phenomena and extend the potential applications, the electronic properties can be easily modulated by the layer number\cite{35,36}, stacking configuration\cite{37,38,39}, mechanical strain\cite{40,41}, sliding\cite{42}, electric and magnetic field\cite{43,44}, chemisorption\cite{45,46,47}, and direct doping\cite{49,50,51} . This paper mainly focuses on the latter two factors. 
\bigskip

How to modulate the fundamental properties becomes one of the main-stream topics in materials science,  chemistry, physics and engineering.  The chemical modification is the most effective method. Pristine graphene has a rather strong $\sigma$ bonding of 2s, 2p\(_x\), and 2p\(_y\) orbitals in a honeycomb lattice. This system creates a quite active chemical environment since each carbon contributes one perpendicular 2p\(_z\) orbital as a dangling bond. That is to say, the host carbon atoms can bond with the various guest atoms (adatoms)\cite{52,53}, molecules\cite{54} and even functional groups\cite{55}. To date, many theoretical\cite{56} and experimental research papers on surface chemisorptions have been published\cite{57}. The previous results show  adsorption-diversified physical and chemical phenomena, such as an opening of the energy gap\cite{58},  semiconductor-metal transitions\cite{59}, the absence/recovery of a Dirac-cone structure\cite{60},  spin-splitting energy bands due to  specific adatoms\cite{61},  diverse van Hove singularities in the density of states (DOS),  single- or multi-orbital hybridizations in C-adatom bonds shown in the spatial charge density and atom- and orbital-decomposed DOSs\cite{62}, and the magnetic moments $\&$ spin configurations\cite{63}. It is well known that  graphitic  systems are the most efficient anode material in  Li$^+$-ion-based batteries. Possibly, the battery efficiency can be enhanced by adding some silicon materials, as observed in the experimental measurements\cite{64}. The Si-atom chemisorption might play an important role in these potential applications. Another chemical modification is  direct doping, the substitution of host carbons by guest adatoms in the honeycomb lattice. For example, the B$_x$C$_y$N$_z$ compounds have been successfully synthesized in the stable structures of 1D nanotubes\cite{65} and 2D layers\cite{66}. Apparently, the essential properties are predicted to present  drastic changes, e.g., in the electronic structures\cite{67}, optical absorption spectra\cite{68} and transport properties\cite{69}. The above-mentioned chemical modifications on graphene systems are worthy of a systematic study on the significant fundamental properties, especially for the critical differences in the orbital hybridizations between the weak and strong Si-C bonds.  
\bigskip

In this paper, a theoretical framework, which is developed within  first-principles calculations, is utilized to fully explore the essential properties of the Si-adsorbed and Si-substituted graphene systems. The concise chemical and physical pictures, the multi-orbital hybridizations without the spin configurations, will be proposed to explain the unusual geometric structures, electronic energy spectra, spatial charge densities, and atom- and orbital-decomposed density of states. That is, all the calculated results are consistent with one another under such mechanisms. Furthermore, they are responsible for the semiconductor-metal transitions after the guest-atom chemisorptions and the creation of energy gaps in the substitution cases. The various experimental characterizations are discussed in detail, and their measurements are required to examine the theoretical predictions, e.g., scanning tunneling microscopy (STM)/transmission electron microscopy (TEM), angle-resolved photoemission spectroscopy (ARPES), and scanning tunneling spectroscopy (STS) experiments, respectively, for C-C $\&$ Si-C bond lengths, low-lying valence bands $\&$ Fermi level, and van Hove singularities in DOSs.

\section{2. Computational methods}

The rich geometric structures and electronic properties of  Si-adsorbed and Si-doped graphene systems are thoroughly explored using the density functional theory (DFT) implemented by Vienna ab initio simulation package (VASP). The many-body exchange and correlation energies, which come from the  electron-electron Coulomb interactions, are calculated from the Perdew-Burke-Ernzerhof (PBE) functional under the generalized gradient approximation. Furthermore, the projector-augmented wave (PAW) pseudopotentials can characterize the intrinsic electron-ion  interactions. As to the complete set of plane waves, the kinetic energy  cutoff is set to be \(\hbar ^2 \left| {k + G} \right|^2 /2m = 500\)
 eV, being suitable for evaluating Bloch wave functions and electronic energy spectra. A vacuum space of \( 10\)  \AA  \ is inserted between periodic images to avoid any significant interaction. The first Brillouin zone is sampled by  \( 9 \times 9 \times 1 \)
   and \( 100 \times 100 \times 1 \) k-point meshes within the Monkhorst-Pack scheme for geometric optimizations and electronic structures, respectively. Such points are sufficient in obtaining the reliable orbital-projected DOSs and spatial charge distributions.  The convergence for the ground-state energy is   
   \(10^{ - 5}   \)
   eV between two consecutive steps, and the maximum Hellman-Feynman force acting on each atom is less than \(  0.01  \)
    eV/\AA \, during the ionic relaxations.      
    
\bigskip    
 Via delicate VASP calculations on certain physical quantities, the critical physical and chemical pictures, i.e., the multi- or single-orbital hybridizations in chemical bonds due to host and guest atoms, can be achieved under a concise scheme. They will be useful in fully comprehending the fundamental physical properties. These important concepts are obtained from the adsorption- and doping-diversified geometric structures, carbon- and silicon-dominated valence and conduction bands, the total charge distributions and their drastic changes after adatom chemisorption or guest-atom doping, and the atom- and orbital-decomposed density of states through  detailed analyses. Also, such physical quantities could shed light on the significant differences between the chemical adsorptions and dopings, such as the metallic or semiconducting behaviors, the normal and irregular electronic energy spectra, and the complicated van Hove singularities, being attributed to the diverse chemical bondings. The developed theoretical framework could  be conceivably generalized to  emergent 2D materials, e.g., the chemical absorption and dopings in layered silicene\cite{70}, germanene\cite{71}, and tinene systems\cite{72}.

\section{3. Results and discussions}
\subsection{3.1. Geometric structures of Si-adsorbed and Si-substituted graphene}

Monolayer graphene has a planar geometry with a honeycomb lattice, being different from the buckled structures in layered silicene\cite{73}, germanene\cite{74}, and tinene\cite{75}. Apparently, this  crystal is  formed by  the very strong \(\sigma\) bonding of 2s, 2p\(_x\), and 2p\(_y\) orbitals, and the weak \(\pi\) bonding of 2p\(_z\) orbitals perpendicular to the graphitic plane. However, the other group-IV systems, with the buckled structures,  are stabilized by the optimal competition between the ${sp^2}$ and ${sp^3}$ chemical bondings. The bond length among all the group-IV systems remains shortest for the C-C (1.42 \AA \ in  Table 1). After the Si-chemisorption on the graphene surface, the bridge site is the most optimal adsorption position among the top and hollow sites, as shown in Fig. 2.  The hexagonal honeycomb lattice of carbon atoms remains a planar structure, while C-C bond lengths are lengthened ${\sim\,1.45}$ \AA \ - ${1.49}$ \AA \ under various adatom concentrations, as shown in Table 1 and Fig. 1(a). Part of the carbon electrons participate in the multi-orbital hybridizations of C-Si bonds, leading to the weakened C-C bondings.  As for the Si-substituted cases, the silicon-carbon honeycomb lattices remain planar structures, indicating sufficiently strong quasi-$\sigma$ bondings due  to the ${sp^2}$-${sp^2}$ multi-orbital hybridizations in the Si-C bonds. The Si-C and C-C bond lengths are, respectively, ${\sim\,1.62}$ \AA \ - ${1.83}$ \AA \ and ${\sim\,1.35}$ \AA \ - ${1.47}$ \AA \ under various substitution cases [Table 1]. The 1:1 substituted system [Fig. 1(b)], the pure silicon-carbon compound, has an optimal Si-C bond length of 1.78 \AA \ ,  much longer than that C-C of 1.42 \AA \  in  pristine graphene. This will lead to a great decrease of the strength in quasi-$\sigma$ bonding,  being further identified by the  largely reduced   charge density in the Si-C bonds. In addition, the guest-atom distribution configurations could be classified into three kinds under specific concentrations lower than 50$\%$ [Figs. 3(a)-(c)], namely, the ortho-, para- and meta-substitution cases. However, the former two are degenerate just at 50$\%$ [Fig. 3(b)].

\bigskip
\begin{table}[htb!]
            \par
           \caption{Energy gap \( E_g\) (eV)/metal(M)/semimetal(SM); C-C bond length (\AA ), Si-C bond length (\AA ), and Si height (\AA ) for Si-substituted/adsorbed graphene.}
                           \begin{center}
                                                
                           \begin{tabular}{llllllllll}
                                \hline Configuration & \makecell{Ratio of \\ Si and C}  & Percentage & \makecell{ \( E_g^{d(i)} \) (eV) \\semimetal(SM) \\metal(M)}  & \makecell{C-C\\(\AA)} & \makecell{Si-C \\ (\AA)}&  \makecell{Si \\height \\(\AA)}  \\ 
                                                \hline
                                            Pristine  & X & X & \( E_g^{d}=0 \) & 1.420 & X & X \\ 
                                        
                                       \multirow{4}{4em}{Adsorption}  & Si:C=6:6 & \(100\%\)  & SM & 1.492  & 2.535 &  2.422 \\

                                            & Si:C=3:6 & \(50\%\) & SM & 1.490 & 2.514 & 2.400\\           
                                           
                                    & Si:C=1:6 & \(16.6\%\) & M & 1.456 & 2.138 & 2.000 \\                       
                                       & Si:C=1:8 & \(12.5\%\) & M & 1.450 & 2.118 & 1.989 \\                   
                                      & Si:C=1:24 & \(4\%\) & M & 1.454 & 2.125 & 1.996 \\                       
                                                     
                                          \hline
                                                   \multirow{4}{4em}{Substitution} & Si:C=3:3 &\(100\%\) & \( E_g^{i}=2.56 \) & X & 1.780 & X  \\  
                                               
                                                   &   
                                                    Si:C=2:4 
                                                   & \(50\%\) ortho &  \( E_g^{d}=0.04 \) & 1.478 & 1.833 & X  \\ 
                                             
                                                   & Si:C=2:4 &\(50\%\) meta & \( E_g^{d}=0.56 \) & 1.465 & 1.829 & X \\   
                                             & Si:C=1:5 &\(20\%\) & \( E_g^{d}=0\) & 1.351 & 1.621 & X \\                
                                      
                                                 & Si:C=1:17 &\(5.8\%\) & \( E_g^{d}=0\) & 1.372 & 1.631 & X \\             
                                   & Si:C=1:23 &\(4.3\%\) & \( E_g^{d}=0\) & 1.379 & 1.649 & X \\                               
                                           
                                          \hline              
                                                                                                 
                                                \end{tabular}
                                                 \end{center}
                                                 \end{table}  

\begin{figure}[tbp!]
\par
\begin{center}
\leavevmode
\includegraphics[width=0.6\linewidth]{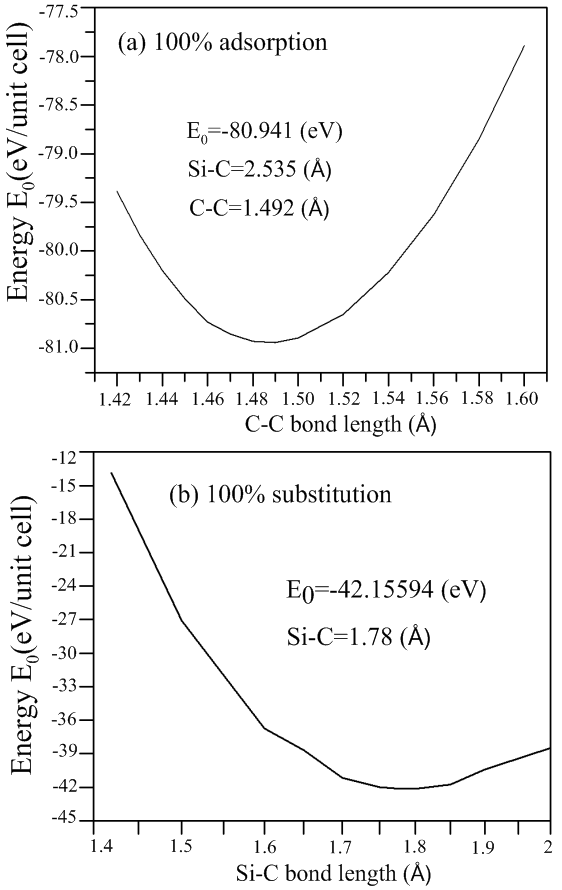}
\end{center}
\par
\caption{Dependence of total ground state energy on C-C and Si-C bond lengths for (a) 100$\%$ double-side adsorption and (b) 100$\%$ substitution, respectively.}
\end{figure}

\begin{figure}[tbp!]
\par
\begin{center}
\leavevmode
\includegraphics[width=0.7\linewidth]{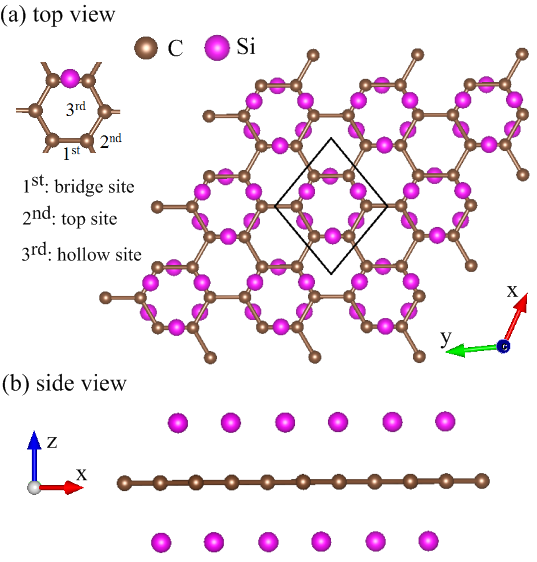}
\end{center}
\par
\caption{Geometric structure of the 100$\%$ Si-adsorbed  graphene: (a) top view and (b) side view.}
\end{figure}

\begin{figure}[tbp!]
\par
\begin{center}
\leavevmode
\includegraphics[width=0.5\linewidth]{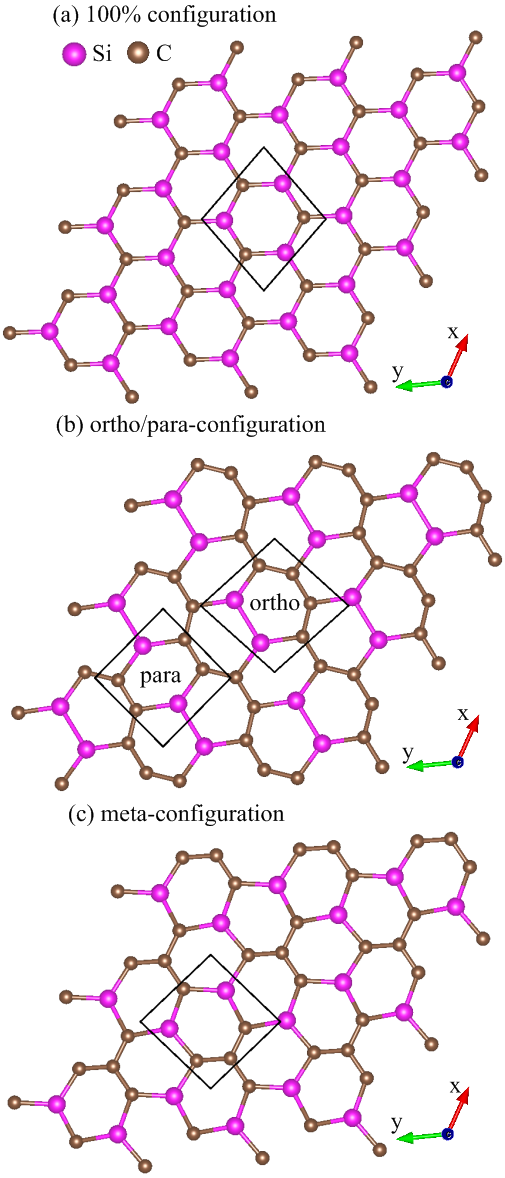}
\end{center}
\par
\caption{Geometric structures of Si-substituted graphene systems: (a) 100$\%$ substitution, (b) 50$\%$ ortho-/para-substitution, and (c) 50$\%$ meta-substitution.}
\end{figure}

\newpage
Scanning tunneling microscopy (STM) is very powerful in identifying the nano-scaled surface morphology\cite{76,77}. This tool can characterize the real-space surface structures under the lateral and vertical atomic resolution, such as the very short bond length, planar or buckled geometries, crystal arrangements, achiral or chiral edges, bridge-, hollow- $\&$ top-site adsorptions, local substitutions, and few-atom clusters. Up to now, a lot of high-resolution STM measurements 
have confirmed the diverse geometric symmetries on the graphene-related systems with the significant ${sp^2}$ chemical bondings, covering the chiral or achiral hexagons on the hollow and cylindrical carbon nanotubes\cite{78}, the various edge structures in nanoscaled-width graphene nanoribbons\cite{79}, the folded, curved, scrolled $\&$ stacked hexagonal lattices\cite{80}, and the AAB, ABC  $\&$ ABA stackings in few-layer graphene systems\cite{81} . The theoretical predictions on  planar graphene after chemisorptions and substitutions, the positions of Si-guest atoms, and Si-C $\&$ C-C bond lengths could all be verified by  STM experiments, being useful in examining the specific  multi-orbital  hybridizations in chemical bonds.
\bigskip

Transmission electron microscopy (TEM) is one of most powerful tools for characterizing  nano-scaled geometric structures, especially with respect to the side-view properties\cite{82}.  A high-energy electron beam, with a very narrow distribution width, can penetrate through an ultra thin sample to generate a diffraction spectrum  by the interactions of charge carriers and condensed-matter systems. To date,  high-resolution TEM measurements can directly identify the side-view crystal structures better than other experimental methods. Plenty of  previous experimental studies show that they have accurately identified the rich and unique geometric properties arising from the carbon-${sp^2}$ bondings in various layered structures,  such as  1D single- $\&$ multi-walled carbon nanotubes\cite{83},  2D curved\cite{84},  folded\cite{85},  scrolled\cite{86},  and stacked  graphene nanoribbons\cite{87}, and the 2D AA-, AB-, ABC- and AAB-stacked few-layer graphene systems\cite{88,89}. The theoretical predictions that cover graphene plane, adatom height and bridge side under the Si-chemisorptions, could be examined by the TEM measurements, indirectly supporting the sp\(^3\)-p multi-orbital hybridizations in Si-C bonds.

\subsection{3.2. Electronic structures}

A pristine monolayer graphene exhibits an unusual band structure, as clearly indicated in Figs. 4(a) and 4(b). The occupied valence bands are asymmetric about the unoccupied ones, mainly 
owing to the multi-orbital $\sigma$ bondings. The low-lying valence and conduction bands, which are initiated from the K and K$^\prime$ valleys (the corners of the first Brillouin zone), are linearly intersecting there. There exist the isotropic Dirac-cone structures at low energy, in which the Fermi level (${E_F=0}$) just crosses at the Dirac point. Apparently, this system belongs to a zero-gap semiconductor because of the vanishing density of state at ${E_F}$. The low-energy bands mainly come from the \(\pi\) bondings of the perpendicular C-${2p_z}$ orbitals. The Dirac-cone structure, being due to the hexagonal symmetry, is predicted to display a lot  of unusual phenomena, e.g.,  diverse magnetic quantizations\cite{90}, Hall effects\cite{91} and optical properties\cite{92}, being consistent with the experimental measurements\cite{93}.  The linear energy dispersions will gradually become  parabolic ones as the state energy of ${|E^{c,v}|}$ grows, or the wave vector deviates from the K/K$^\prime$ point. Specifically, the  middle-energy parabolic valence and conduction bands present the saddle points at M point, respectively, corresponding to ${\sim-2.4}$ eV and 1.8 eV. Such critical points in the energy-wave-vector space could be regarded as the significant band-edge states in creating the important van Hove singularities. That is to say, they are thus expected to induce  special structures in the essential physical properties. The $\sigma$ valence bands at the deeper energy come into existence at ${E^v\sim\,-3}$ eV from the $\Gamma$ point, being regarded as the extreme point of parabolic energy dispersion. Their electronic states are formed by the 2s, 2p\(_x\), and 2p\(_y\) orbitals of carbon atoms, or the very strong \(\sigma\) bondings on the graphene plane.  
\bigskip

The band structure is dramatically changed by the Si-chemisorptions. The band asymmetry about the Fermi level becomes more obvious.  For the \(100\%\) double-side adsorption, as shown in Fig. 4(c), there exist some valence and conduction bands simultaneously intersecting with the Fermi level (${E_F=0}$), so this system exhibits  semi-metallic behavior.  Apparently, the distorted Dirac-cone structure appears near the $\Gamma$ point. The separation of valence and conduction Dirac points could reach ${\sim\,0.5}$ eV. The electronic energy spectrum is highly anisotropic energy one along $\Gamma$M and $\Gamma$K. The occupied electronic states come to exit between the Fermi level and the bottom of the conduction-band states. This clearly indicates the creation of free conduction electrons by the effective adatom dopings. On the other hand,  free holes are generated in the  unoccupied valence states along M$\Gamma$ and K$\Gamma$ lines. As a result, it is difficult to identify Si-adsorbed graphene as a n-type or p-type system. However, this system belongs to a 2D semimetal, since it has a finite density of states at the Fermi level (Fig. 8(b)) arising from the crossing valence and conduction subbands. The above-mentioned unusual band structures are closely related to the very strong competition/cooperation of orbital hybridizations in Si-C and C-C bonds. Also, some drastic changes in electronic structures are revealed in the \(100\%\) single-side adsorption case, as clearly indicated by a comparison of Fig. 4(d) and Fig. 4(c). The chemisorption of silicon adatoms induces the free electrons and holes simultaneously,  similar to the  double-side case (Fig. 4(c)). The low-lying conduction bands near the $\Gamma$ point and the vacant valence bands along M$\Gamma$ and K$\Gamma$ lines reduced in number. That is to say, the number of energy bands intersecting with the Fermi level declines for a further decrease of the Si-concentration, and the 2D free carrier density behaves so. As to both adsorption cases, carbon host atoms and  silicon guest ones make significant contributions to the electronic structures of the whole energy range, in which their dominances are obviously displayed by the red triangles and blue circles. These important results mean that there exist  non-negligible multi-orbital hybridizations in Si-C, C-C and  Si-Si bonds.  Furthermore, there is an obvious difference from the aforementioned \(100\%\) adsorption cases when the concentration decreases [Figs. 4(e) and 4(f)]. That is, only the valence energy bands intersect with the Fermi level. This indicates that unoccupied electronic states between the Fermi level and the top of the valence-band states all belong to free holes. As a result, these systems can be regarded as the p-type metal. Especially, the anisotropic Dirac cone structure without/with separation [Fig. 4(e)/(f)] between the $\Gamma$ and K points appears below the Fermi level.  Such unusual valence Dirac cone structure is responsible for the deformed V-shape structure in Fig. 8(d) [discussed in the density of states section]. 

\bigskip
The atom- and orbital-dominated energy bands are worthy of a closer examination. For 100\(\% \) double-side and single-side adsorption cases, most of energy bands are co-contributed by C-host and Si-guest atoms, with part of them mainly coming from either the former or the latter. In general, the low-lying and middle-energy valence and conduction bands are dominated by the Si-guest atoms. The percentage of atom contribution is about \(4:1\) (\(2:1\)) in the Si-100\(\% \) (Si-50\(\% \)) adsorption case, as shown in Fig. 4(c) (Fig. 4(d)), being estimated from the  \(sp^3  -p\)  bonding in the Si-C bond [the conclusion from the atom- and orbital-projected density of states; discussed later in Figs. 8(b) and 8(c)]. The four (3s, 3p\(_x\), 3p\(_y\), 3p\(_z\)) orbitals of silicon and the single 2p\(_z\) orbital of carbon make important contributions to such energy bands. Apparently, there are few $\sigma$ valence bands of (2p\(_x\), 2p\(_y\)) orbitals near the $\Gamma$ point;  they belong to the concave-downward energy dispersions at ${E^v\sim\,-4.1}$ eV \(\&\) $-4.2$ eV for the 100\(\% \) adsorption cases [Figs. 4(d) and 4(c)], as identified from Figs. 8(c) and 8(b). It should be noted that the pristine $\sigma$ valence bands come into existence at ${E^v\sim\,-3.0}$ eV [Figs. 4(a,b) and 8(a)]. As the concentration declines, Si adatom- and C atom-co-dominated  energy bands only appear at specific energies, while other bands are mainly dominated by C atoms, as indicated in Figs. 4(e) and 4(f) for \(16.6\%\) and \(12.5\%\) adsorption cases, respectively. Also, the $\sigma$ valence bands of (2p\(_x\), 2p\(_y\)) orbitals situate at ${E^v\sim\,-3.5}$ eV and ${E^v\sim\,-3.2}$ eV [Figs. 4(e) and 4(f)]. This clearly illustrates that the shorter C-C bond lengths under low-concentration adsorptions, as compared with the 100\(\% \) cases [Figs. 4(c,d)].

\bigskip
Both silicon and alkali adatoms can create  free carriers, while their band properties are quite different from each other. The alkali-adsorbed graphene systems present  approximately rigid energy bands and a few Li-dominated conduction bands, clearly indicating the blue shift of the Fermi level\cite{62}. Their free carriers purely originate from the electron charge transfer from the outermost s orbital of each alkali adatom to the carbon host atom. Furthermore, the atom- and orbital-projected densities of states clearly show the weak, but significant s-p\(_z\) orbital hybridization in every alkali-carbon bond. On the other hand, the free electrons and holes in the  Si-adsorbed cases are associated with a strong overlap of valence and conduction bands, so the shift of $E_F$ is very difficult to characterize in value, except for the low-concentration adsorptions, in which the Fermi level exhibits a red shift. There are a lot of extra Si-dominated and (Si, C)-co-dominated energy bands in the whole energy spectrum, especially for those crossing at the Fermi level. The complicated chemical bondings are deduced to survive in Si-adsorbed graphene systems, in which they cover the 2p\(_z\)-(3s, 3p\(_x\), 3p\(_y\), 3p\(_z\)),  (3s, 3p\(_x\), 3p\(_y\), 3p\(_z\))-(3s, 3p\(_x\), 3p\(_y\), 3p\(_z\)) and (2s, 2p\(_x\), 2p\(_y\))-(2s, 2p\(_x\), 2p\(_y\)) multi-orbital hybridizations in the C-Si, Si-Si and C-C bonds. Most importantly, the above-mentioned differences obviously illustrate the adatom-adsorption-induced diverse phenomena and the critical mechanisms/pictures in determining the fundamental properties.

\bigskip
Electronic structures obviously show drastic changes in the presence of Si-substitutions, as clearly indicated in Figs. 5(a-f). The asymmetry of valence and conduction bands about \(E_F=0\) is greatly enhanced after various substitutions, e.g., 100\(\%\) case in Fig. 2-5(a), 50\(\%\) case in Figs. 5(b) and 5(c), 20\(\%\) case in Fig. 5(d), 5.8\(\%\) case in Fig. 5(e), and 4.3\(\%\) case in Fig. 5(f).  The substitution and adsorption cases sharply contrast with each other  in Figs. 5 and 4. First, all the substitution configurations and concentrations show the semiconducting behavior with a finite or vanishing band gap. For example, the 100\(\%\) Si-substituted graphene is a wide-gap semiconductor with an energy gap of \( E_g^{i}=2.56 \) eV. Energy gaps are direct or indirect being determined by the highest valence and the lowest conduction state near the $\Gamma$ point. Their values decline with decreasing of the Si-concentrations and are strongly depended on the guest-atom distribution configurations, e.g., \( E_g^{d}=0.56 \) eV  for the meta-configuration [Fig. 5(c)] under the 50\(\%\) substitution.  Also,  the zero-gap semiconducting behavior is revealed in the  50\(\%\) ortho-case [Fig. 5(b)], where an anisotropic Dirac-cone structure appears at a certain  ${\bf k}$-point between $\Gamma$ and K.  Only one Dirac point intersects with the Fermi level, so its density of states vanishes there [Fig. 9(b)]. This is responsible for the zero-gap semiconductor. When the substitution concentration declines, the unusual zero-gap semiconducting behavior appears, in which only one conduction Dirac point intersects with the Fermi level near the $\Gamma$ point, as shown in Figs. 5(d-f).  Second, the linear Dirac cone at $\Gamma$ point in Fig. 4(a,b) is seriously separated and distorted in Fig. 5(b), or even thoroughly destroyed in Figs. 5(a) and 5(c). Also, this Dirac cone structure is  significant deformed without separation and slightly shifted to the valence band at  $\Gamma$ point, as shown in Figs. 5(d-f).  Third, the valence and conduction bands near the Fermi level are initiated from the $\Gamma$ point, but almost independent of the M and K points.  Finally, all the energy bands are mainly co-dominated by the Si-guest and carbon-host atoms under the high substitutions [Figs. 5(a-c)]. However, these Si-guest and C-host atoms-co-dominated energy bands only exist at certain energies under the lower substitutions [Figs. 5(d-f)]. Furthermore, the separated $\sigma$ bands at deep energies purely due to C-(2p\(_x\), 2p\(_y\)) orbitals are absent. These results suggest the existence of  quasi-$\sigma$ and quasi-$\pi$ bondings, respectively, originating from the (3s, 3p\(_x\), 3p\(_y\))-(2s, 2p\(_x\), 2p\(_y\)) $\&$ 3p\(_z\)-2p\(_z\) orbital hybridizations [supported by the spatial charge distributions in Fig. 7 and the atom- and orbital-projected density of states in Fig. 9].

\begin{figure}[tbp]
\par
\begin{center}
\leavevmode
\includegraphics[width=0.5\linewidth]{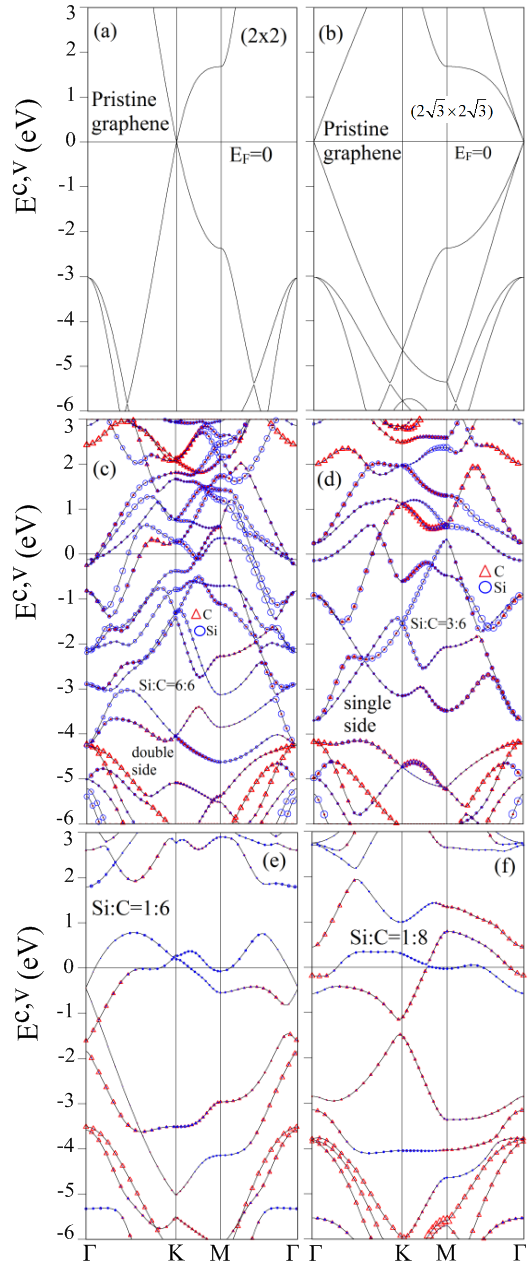}
\end{center}
\par
\caption{Electronic structures with the dominances of host and guest atoms for Si-adsorbed graphene systems: (a) pristine case, (b) pristine graphene under the enlarged unit cell identical to the 100$\%$ case, (c) \(100\%\) double-side adsorption, and (d) \(100\%\) single-side adsorption,  (e) \(16.6\%\) adsorption, and (f) \(12.5\%\) adsorption.}
\end{figure}

\begin{figure}[tbp]
\par
\begin{center}
\leavevmode
\includegraphics[width=0.5\linewidth]{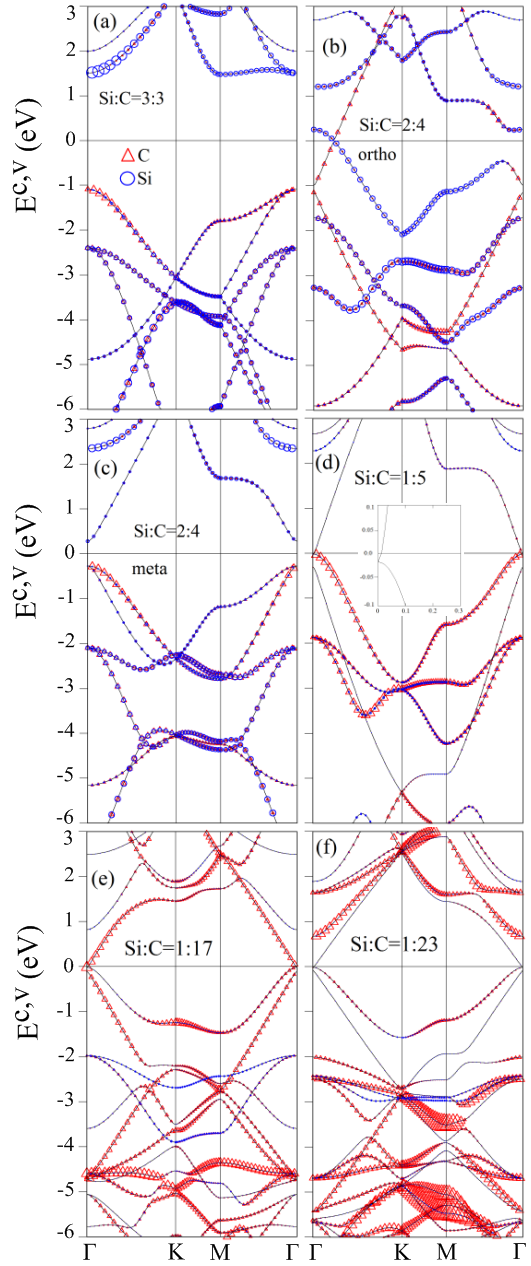}
\end{center}
\par
\caption{C- and Si-dominated valence and conduction bands of  Si-substituted graphene systems: (a) 100$\%$ substitution, (b) 50$\%$ ortho-substitution, (c) 50$\%$ meta-substitution, (d) 20$\%$ substitution, (e) 5.8$\%$ substitution, (f) 4.3$\%$ substitution.}
\end{figure}

\newpage
Angle-resolved photoemission spectroscopy (ARPES) is the only tool for measuring the wave-vector-dependent occupied electronic states, especially for valence and conduction bands crossing at the Fermi level. Details of the experimental equipments can be  found in the books written by Tran., et al\cite{93} and Lin., et al\cite{94}. To date,  high-resolution ARPES measurements have identified the rich band structures in graphene-related systems, being greatly diversified by the distinct geometric symmetries, dimensions, stacking configurations,  numbers of layers, and adsorptions, substitutions $\&$ intercalations. For example, the verified electronic structures, which are consistent with the theoretical predictions\cite{95} and the experimental measurements\cite{96}, include parabolic valence bands with an energy gap in the 1D graphene nanoribbons\cite{97}, the linear Dirac-cone structure in monolayer graphene\cite{98}, the blue shift of the Fermi level in alkali-adsorbed graphene systems\cite{99}, two parabolic valence bands near $E_F$ without an energy gap in bilayer AB-stacked graphene\cite{100}, the monolayer- and bilayer-like valence bands in trilayer ABA stacking, the partially flat, sombrero-shaped, and linear bands in trilayer ABC stacking\cite{101}, and the parabolic $\&$ linear energy dispersions near the K and H points (${k_z=0}$ and $\pi$) in a natural graphite\cite{102}.  Similar ARPES measurements are available for thoroughly examining the significant effects on band structures after the chemisorption and substitution of Si adatoms. They are conducted on the finite or vanishing band gap, the numbers of the valence and conduction bands intersecting with the Fermi level near the $\Gamma$, K and M points, or only the valence energy bands intersecting with the Fermi level, and the existence/destruction of the carbon-dominated $\sigma$ bands at the $\Gamma$ point initiated from ${-4.1}$ eV,  ${-4.2}$ eV, and ${-3.5}$ eV. Such detailed information is sufficient to determine the critical multi-orbital hybridizations of C-Si bonds, the p-sp\(^3\) or sp\(^2\)-sp\(^2\) bondings.

\section{3.3. Spatial charge densities}

The multi-orbital hybridizations in chemical bonds, which are responsible for the rich geometric structures, energy bands and density of states, could be delicately identified from the spatial charge densities (${\rho}$'s) and their variations (${\Delta\rho}$'s) after  various modifications. The latter is obtained from the difference between the Si-chemisorption/Si-substitution and pristine cases. A pristine graphene, as clearly shown in Fig. 6(a), presents a very high carrier density between two carbon atoms [red region enclosed by a black rectangle], indicating a rather strong $\sigma$ bonding due to three C-(2s, 2p\(_x\), 2p\(_y\)) orbitals on the honeycomb lattice. Such bonding is hardly affected by the Si-adatom adsorptions [Figs. 6(b), 6(d), and 6(f)]. Also, there exists the $\pi$ bonding near the plane boundary along the $z$-direction [area covered by a red rectangle]. The 2p\(_z\)-2p\(_z\) orbital hybridizations in C-C bonds  might be drastically changed under the Si-chemisorptions. The charge distributions related to silicon adatoms and carbon atoms along the $x$-, $y$-, and $z$-directions present  obvious variations. The strong evidences are illustrated by ${\Delta\rho}$'s in Figs. 6(c), 6(e), and 6(g). The charge density is enhanced near the carbon atoms on the ${(z, x)}$- and ${(z, y)}$-planes [red regions]. This result means that some electronic charges transferred from Si adatoms to C atoms. In addition to the $z$-direction, the important charge variations along the $x$- and $y-$directions  survive between silicon adatoms and carbon atoms/silicon ones, indicating the multi-orbital hybridizations in Si-C and Si-Si bonds. According to the direction- and  position-dependent variations of
charge densities, there exist (3s, 3p\(_x\), 3p\(_y\), 3p\(_z\))-2p\(_z\) and (3s, 3p\(_x\), 3p\(_y\), 3p\(_z\))-(3s, 3p\(_x\), 3p\(_y\), 3p\(_z\)) complicated  interactions in Si-C and Si-Si bonds, respectively. 

\begin{figure}[tbp]
\par
\begin{center}
\leavevmode
\includegraphics[width=0.7\linewidth]{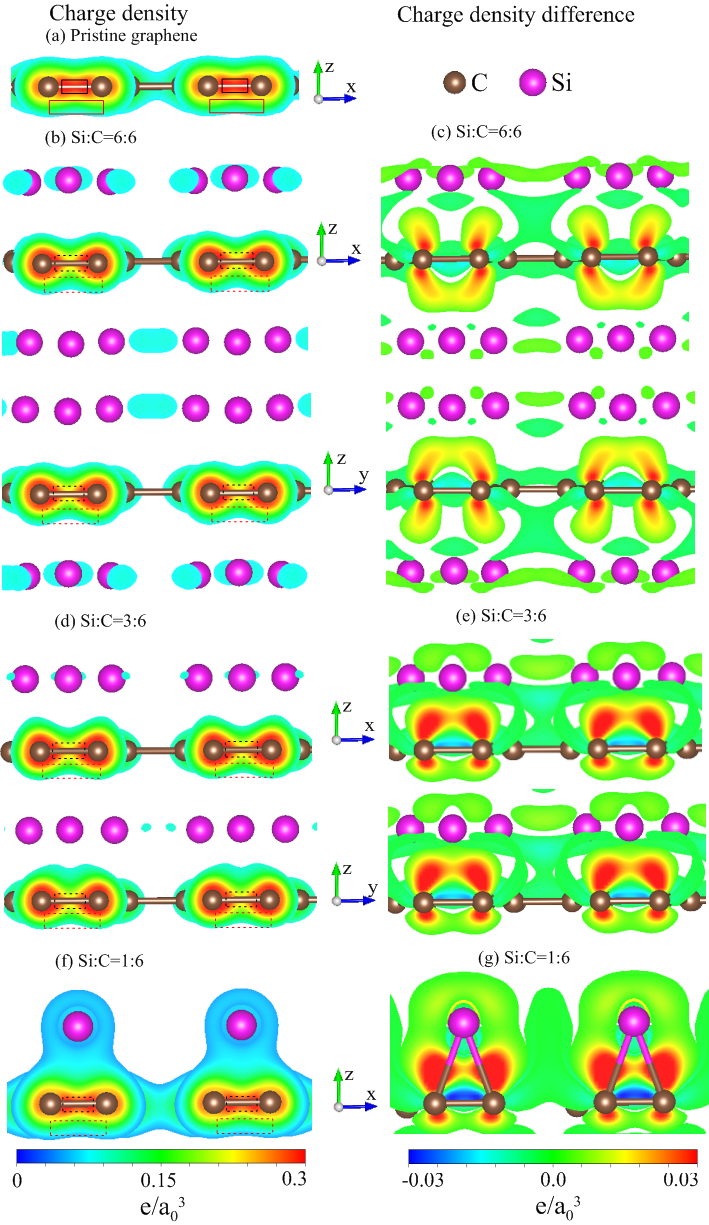}
\end{center}
\par
\caption{Spatial charge density/its variation after chemisorption for (a) pristine system, (b)/(c) 100$\%$ double-side adsorption, (d)/(e)  100$\%$ single-side adsorption, and  (f)/(g) \(16.6\%\) adsorption.}
\end{figure}

\newpage

The Si-substitution cases exhibit diversified charge densities, compared with the pristine and Si-chemisorption ones. For the 100\(\%\) substitution, the Si-C bonds, which form a honeycomb lattice [Fig. 3(a)], present sufficiently high charge densities between two neighboring atoms [Fig. 7(a)]. They are formed by the quasi-$\sigma$ bondings, in which the charge-density-dependent strengths are greatly reduced, compared to the $\sigma$ ones in C-C bonds [Fig. 6(a)]. This  is consistent with the longer Si-C bonds and the shorter C-C bonds. Similar results are revealed in other substitution cases, e.g.,  ${50}$$\%$ substitution under the ortho-, and meta-configurations [Figs. 7(c) and 7(e)], and ${20}$$\%$ substitution in Fig. 7(g). There also exist the quasi-$\pi$ bondings near the boundary. These present the  non-well-behaved charge distributions, compared with  those of a pristine graphene [Fig. 6(a)]. The obvious variations of charge densities on the ${(z, y)}$- and ${(z, x)}$-planes, being clearly show in Figs. 7(b), 7(d), 7(f),  and 7(h), suggest  significant
(3s, 3p\(_x\), 3p\(_y\))-(2s, 2p\(_x\), 2p\(_y\)) $\&$ 3p\(_z\)-2p\(_z\) orbital hybridizations in Si-C bonds. The coexistence of  multi- and single-orbital interactions are further supported by the atom- and orbital-decomposed densities of states [discussed later in Figs. 9]. 

\begin{figure}[tbp]
\par
\begin{center}
\leavevmode
\includegraphics[width=0.7\linewidth]{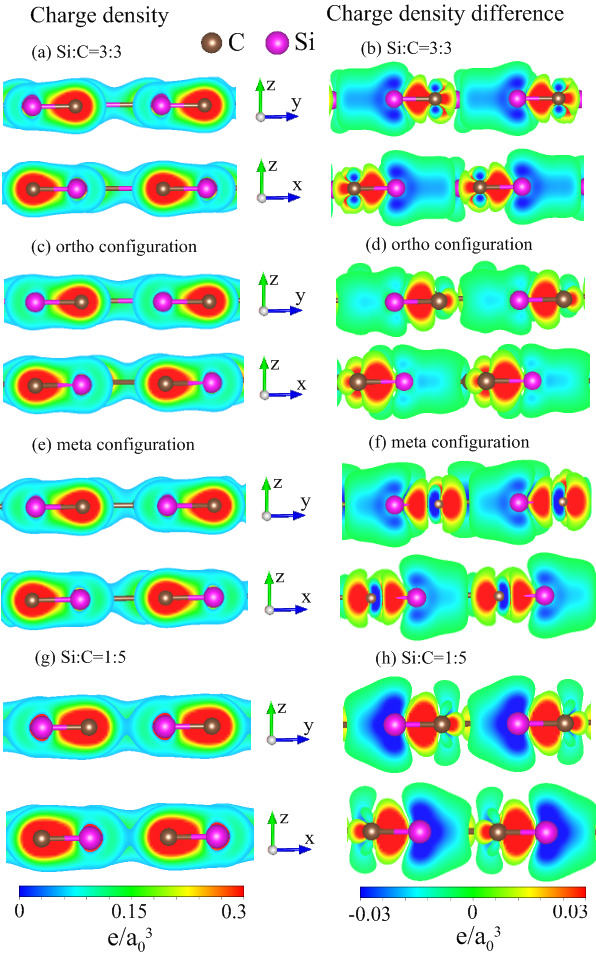}
\end{center}
\par
\caption{Similar plot as Fig. 6, but shown for Si-substituted graphene systems for (a)/(b) 100$\%$ substitution, (c)/(d)  50$\%$ ortho-substitution,  (e)/(f) 50$\%$ meta-substitution, and (g)/(h)  20$\%$ substitution.}
\end{figure}

\section{3.4 Orbital-projected density of states}

The main features of energy dispersions are directly reflected in the diversified density of states. Special structures, the van Hove singularities, mainly originate from the critical points in the energy-wave-vector space, in which the band-edge states might belong to the local minima $\&$ maxima, and the saddle points. In general, there are three kinds of novel structures, the V-shaped structure crossing at the Fermi level, logarithmically divergent peaks [${E\sim\,}$${-2.4}$ eV $\&$ 1.8 eV] and shoulders [${E\sim\,-3}$ eV], as clearly observed in a  pristine graphene [Fig. 8(a)]. They are, respectively, due to the linear Dirac cone at the K/$\Gamma$ point in Fig. 4(a)/(b), the saddle points at the M points, and the extreme points of parabolic dispersions at the $\Gamma$ point. Specifically, the former two structures are generated by the $\pi$ bonding of \(2p_z\) orbitals (dashed red curve), whereas the initial $\sigma$ bands are closely related to the (2p\(_x \), 2p\(_y \)) orbitals [dashed green and blue curves]. The $\pi$-bonding structures are separated from those of the $\sigma$ bondings. 

\bigskip

Apparently, a lot of van Hove singularities in the density of states are created by the Si-adatom chemisorption in the \(100\%\) double- and single-side cases, being obviously displayed in Figs. 8(b) and 8(c), respectively. The finite density of states at the Fermi level indicate the semi-metallic behavior, while the vanishing value in the pristine case [Fig. 8(a)] corresponds to a zero-gap semiconductor. The chemical bonding between Si and C atoms is responsible for the significant overlap of the valence and conduction bands and the creation of new energy bands [Figs. 4(c) and 4(d)]. The significant  contributions of carbon-2p\(_z\) orbitals appear in the whole energy range of ${-6}$ eV ${\le\,E\le\,3}$ eV. Specifically, for that above ${-4}$ eV, the non-negligible contributions from the four Si-(3s, 3p\(_x\), 3p\(_y\), 3p\(_z\)) orbitals come into existence, respectively, indicated by the solid pink, green, orange and purple curves in Figs 8(b) and 8(c). The van Hove singularities from these orbitals are merged. This clearly indicates the p-sp\(^3\) orbital hybridization in C-Si bonds. The multi-orbital hybridizations, which replace the  2p\(_z\)-orbital bondings on the graphene plane, are also confirmed by the previous charge density distributions [Figs. 6(b)-(g)]. However, the $\sigma$ bonding of carbon atoms are hardly affected by the Si adsorptions, in which they are represented by the isolated shoulder structure below ${-4.2}$ eV (${-4.1}$ eV) in the Si-100$\%$ (Si-50\(\%\))  adsorption cases. When the concentration decreases, as shown in Fig. 8(d), only a few of van Hove singularities at certain energies are created by Si-guest atoms. The seriously deformed valence V-shaped structure and two logarithmic divergent peaks at \(-3\) eV and \(-0.5\) eV come to exist, respectively, resulting from the significant distorted valence Dirac cone structure and two saddle points at the M point in Fig. 4(e).  The finite density of states at the Fermi level also exist; however, it is responsible for the metallic behavior. Furthermore, the $\sigma$ shoulder structure of C-(2p\(_x\) and 2p\(_y\)) orbitals appears at \(-3.5\) eV, indicating the stronger $\sigma$ C-C bonds (shorter C-C bond lengths) in the low-concentration system compared with the \(100\%\) adsorption cases. 

\bigskip
Why are valence and conduction bands in Fig. 4 dramatically altered by the Si-adsorbed graphene? The diversified band properties are clearly identified from three critical mechanisms. First, the chemical modification belongs to the significant surface adsorption, but not the stronger substitution. This means that $\sigma$ bondings in planar C-C bonds exist even after  significant Si-chemisorptions. As a result, the $\sigma$ bands of the C-(2p\(_x\), 2p\(_y\)) orbitals exhibit a significant red shift of ${\sim\, 1}$ eV [red triangles at ${E^v\sim\,-4.1}$ eV and ${E^v\sim\,-4.2}$ eV in Figs. 4(d) and 4(c)] under \(100\%\) adsorption cases, and this red shift is smaller ${\sim\, 0.5}$ eV [Fig. 4(e)] in the low-concentration case. The red shift value is determined by the ionization-energy difference between carbon and silicon atoms.  Second, the important sp\(^3\)-p multi-orbital hybridizations are covalently formed by the four half-occupied orbitals of (3s, 3p\(_x\), 3p\(_y\), 3p\(_z\)) in Si adatoms and one similar orbital of ${2p_z}$ in the C host atoms. For example,  plenty of low-lying valence and conduction bands are dominated by the silicon guest atoms. Furthermore, their contributions to the total density of states are higher than those due to the carbon atoms except in the energy range below ${-4}$ eV,  respectively, indicated by the solid and dashed curves Figs. 8(b) and 8(c). Finally, from the viewpoint of the separated atomic orbital energies, all the atomic orbitals in the  silicon adatoms have larger spatial distributions and  smaller binding energies, compared with those of the carbon atoms. These orbitals, which take part in the Si-C bonds, create many energy bands in the available energy range of the electronic energy spectrum, especially for the valence and conduction bands intersecting with the Fermi level. Apparently, the Dirac-cone band structure without any valence and conduction overlap, is thoroughly replaced by the Si-dominated electronic structures under  full chemisorptions. The other essential physical properties are expected to show diverse phenomena under the Si-chemisorptions, e.g., magnetic quantizations\cite{89}, optical absorption spectra\cite{91}, and transport properties\cite{69}.

\begin{figure}[tbp!]
\par
\begin{center}
\leavevmode
\includegraphics[width=0.9\linewidth]{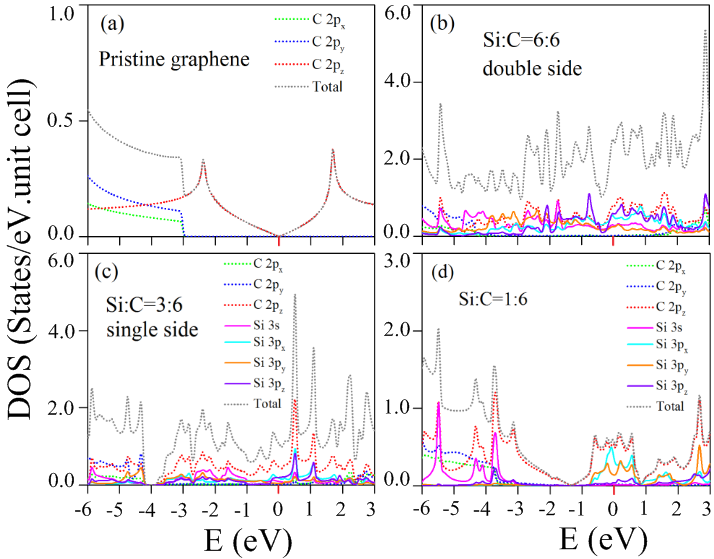}
\end{center}
\par
\caption{Orbital-projected density of states for Si-adsorbed graphene systems: (a) pristine case, (b) 100$\%$ double-side adsorption, (c) 100$\%$ single-side adsorption, and (d) 16.6$\%$  adsorption.}
\end{figure}

\newpage
 The substitution and adsorption cases are thoroughly different from each other in the main feature of density of states. For the former chemical modifications, the number of electronic states, which is revealed in Fig. 9, vanishes within the specific band-gap region centered at the Fermi level.  Most of the substitution configurations and concentrations correspond to the finite gap semiconductors, e.g., energy gaps due to the highest occupied valence state and the lowest unoccupied conduction one at the $\Gamma$ point under the 100$\%$ substitution [Fig. 5(a)] and  meta-50$\%$ substitution [Fig. 5(c)]. Only the ortho-50$\%$ case in Fig. 5(b) belongs to the zero-gap semiconductors with a seriously distorted Dirac-cone structure between the $\Gamma$ and K points. In the case of the lower concentration substitutions, the unusual zero-gap semiconducting behavior and the distorted valence Dirac cone structure [Fig. 5(d)], respectively,  resulting in the almost negligible density of states at the Fermi level and the seriously deformed V-shape structure in Fig. 9(d).  The van Hove singularities can create many various special structures, namely, the obvious V-shape structure across ${E_F}$, strong shoulders/asymmetric peaks, and prominent symmetric peaks, as observed in the chemisorption cases [Fig. 8]. Apparently, the atom- and orbital-projected densities of states show that the contributions coming from the Si-3p\(_z\)  and C-2p\(_z\) orbitals [solid purple and dashed red curves in Figs. 9(a-d)] appear simultaneously. Furthermore, the merged special structures are also revealed in the other orbitals, e.g.,  the interactions of Si-(3s, 3p\(_x\), 3p\(_y\))  and C-(2s, 2p\(_x\), 2p\(_y\))  orbitals. These results clearly illustrate the multi-orbital hybridizations of sp\(^2\)-sp\(^2\) $\&$ p-p in Si-C bonds. That is to say, the Si-C bonds present  quasi-$\sigma$ and quasi-$\pi$ chemical bondings.  In addition, it should have the $\sigma$ and $\pi$ bondings in C-C bonds.  The predicted orbital hybridizations are consistent with the spatial charge distributions in Fig. 7 and can account for the  Si-substitution-enriched band structures in Fig. 5, respectively.
 
 \begin{figure}[tbp!]
 \par
 \begin{center}
 \leavevmode
 \includegraphics[width=0.9\linewidth]{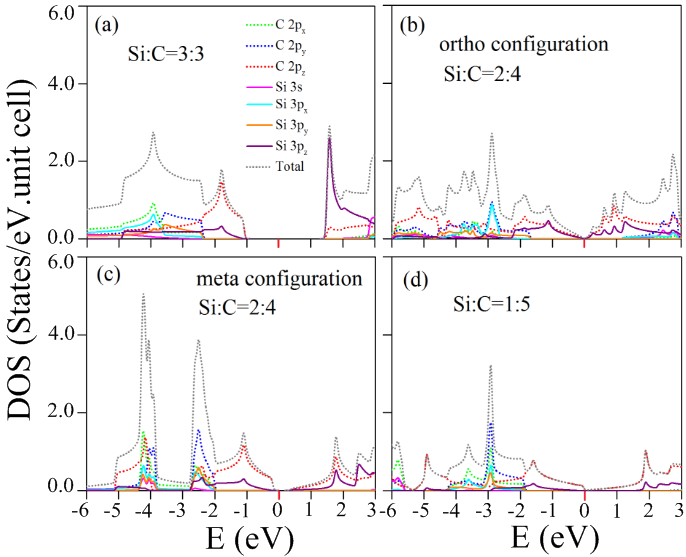}
 \end{center}
 \par
\caption{Similar plot as Fig. 8, but displayed for Si-substituted graphene systems: (a) 100$\%$ substitution, (b) 50$\%$ ortho-substitution, (c) 50$\%$ meta-substitution, and (d) 20$\%$ substitution.}
 \end{figure}
 
\newpage
Scanning tunneling spectroscopy (STS) can provide sufficient information on the density of states at the Fermi level and the various van Hove singularities due to the valence and conduction bands simultaneously. High-resolution STS measurements are available for distinguishing the semiconducting and metallic behavior. Furthermore, they are very useful in identifying the close relations between the electronic energy spectra and the orbital hybridizations of the significant chemical bonds. Such experimental measurements have been successfully utilized to verify the band properties near the Fermi level and the dimension-diversified van Hove singularities in the graphene-related systems even in the presence of magnetic field, such as 2D few-layer graphene systems with the AB, ABC, AAB stackings\cite{103,104}, 1D metallic and semiconducting carbon nanotubes\cite{105,106}, and 3D Bernal graphite\cite{107}. Apparently, the theoretical predictions on the Si-adsorption- and Si-substitution-diversified density of states in monolayer graphene systems, which could be examined by  the STS experiments, cover the finite or vanishing density of states at the Fermi level, and the low- and middle-energy van Hove singularities. That is, such experimental examinations can provide the critical information on the multi-orbital hybridizations in the p-sp\(^3\) or sp\(^2\)-sp\(^2\) chemical bondings.

\bigskip
The essential properties of Si-adsorbed graphene systems might sharply contrast with other adatom-decorated ones. For example, graphene oxides have a bridge-site adsorption position\cite{93}, as observed in the Si-adsorption case. However, the former present slight buckling and a shorter adatom-C bond length. All graphene oxides belong to wide-, narrow- or zero-gap semiconductors, without any band-overlap-induced free electrons and holes. The O-${2s}$ orbitals do not take part in the low-lying energy bands, while the opposite is true for the Si-${3s}$ ones. Under high adatom concentrations (\(\ge 50\% \)), valence and conduction bands in the range of \(- 1.5  \le E^{c,v}  \le 1.5\) eV are absent for graphene oxides, being thoroughly different from the frequent crossing over the Fermi level in the current case. Apparently, the \(\sigma \) bands purely due to the carbon atoms disappear. Moreover, the three (2p\(_x\), 2p\(_y\), 2p\(_z\)) orbitals of oxygen and carbon atoms would dominate the multi-orbital hybridizations in O-C, O-O and C-C bonds. On the other side, the 2p\(_z\) and (2p\(_x\), 2p\(_y\)) orbitals of carbon atoms, respectively, make important contributions to C-C and Si-C bonds; furthermore, the Si-(3s, 3p\(_x\), 3p\(_y\), 3p\(_z\)) orbitals participate in the Si-C and Si-Si bonds. The above-mentioned important differences could be verified by high-resolution STM,  TEM, ARPES and STS measurements.

\section{4. Conclusion}

 The essential properties of Si-modified graphene systems have been explored by the first-principles method in detail. Apparently, the geometric structures, band structures, spatial charge densities, and atom- and orbital-projected density of states exhibit  rich and unique phenomena, being sensitive to the chemisorption $\&$ substitution cases, and the concentration $\&$ distribution configuration of the Si-guest atoms.  The complex multi-orbital hybridizations/significant chemical bondings are proposed to explain the Si-induced chemical modifications on graphene systems. The calculated results clearly show that  free carriers and an energy gap might be created by the chemical adsorptions and substitutions, respectively. The distinct modification ways of silicon guest atoms can modulate the semiconducting, metallic, and semi-metallic behavior, in a way difficult to  observe in other kinds of atoms\cite{108}. For example, there are certain important differences between silicon and alkali atoms\cite{109} in terms of chemisorptions, e.g., the multi-orbital or single-orbital hybridizations, the existence or absence of a charge transfer, a blue or undefined shift of the Fermi level, only  conduction electrons or free conduction electrons $\&$ holes, and the dramatic or slight changes for the low-lying energy bands. The Si-adsorbed and Si-substituted  graphenes could serve as  potential candidates in nanoscaled-applications\cite{110,111}. A similar theoretical framework could  be generalized for emergent layered  materials, such as chemical adsorptions and substitutions on silicene\cite{112}, germanene\cite{113}, stanene\cite{114}, phosphorene\cite{115}, and bismuthene\cite{116}.
 
 \bigskip
 The chemisorptions and substitutions of Si-guest atoms on monolayer graphene present unusual geometric properties, being directly revealed in the spatial charge distributions. Si-adsorbed graphene is a non-buckled plane, in which the optimal position corresponds to the bridge site. A planar structure clearly indicates a very small variation in the $\sigma$ bonding of C-(2s, 2p\(_x\), 2p\(_y\)) orbitals after the Si-chemisorptions; that is, these three orbitals do not take part in Si-C bonds.  This adsorption configuration is similar to monolayer graphene oxide [93], while there is a slight buckling in the latter. The Si-C bond length is ${\sim\,2.1}$ \AA   - ${2.5}$ \AA,  suitable for the significant multi-orbital hybridizations of Si-C bonds. Furthermore, the C-C bond lengths display a minor change. Via detailed analyses on the 3D charge density, its spatial variation, and the atom- and orbital-projected density of states, the sp\(^3\)-p multi-orbital hybridizations are deduced to dominate the chemical Si-C bonds. Apparently, the three physical quantities are consistent with one another. The theoretical predictions of the bridge site and C-C bond length (Si height) could be examined by high-resolution STM (TEM) measurements. Concerning Si-substituted graphene systems, they remain in the planar structure, as observed under the chemisorption cases. There exist  very strong multi-orbital chemical bondings between Si and C atoms.  The Si-C bond length is about 1.62 \AA \ - 1.83 \AA, being shorter than those in the Si-adsorbed systems. This means that more C-orbitals are strongly hybridized with the four Si-orbitals. STM experiments are available for examining the predicted Si-C and C-C bond lengths. 
 
\bigskip
 Any chemisorptions and substitutions of Si-guest stoms thoroughly alter the unusual band structure of monolayer graphene, especially for the zero-gap semiconducting behavior and linear Dirac cone due to the $\pi$ bondings of the 2p\(_z\) orbitals. In the \(100\%\) double- and  single-side adsorption cases, the Si-adsorbed graphene systems are semi-metals with free conduction electrons and valence holes. For the low-concentration systems, they belong to the p-type metals with only free valence holes. The Dirac-cone structure near the $\Gamma$ point is seriously distorted after the Si-chemisorptions. There are more valence and conduction bands, accompanied by  various band-edge states, in the whole electronic energy spectrum, e.g., the emergent low-lying energy bands along the K$\Gamma$ and M$\Gamma$ directions. However, the $\sigma$ bands, which arises from the (2p\(_x\), 2p\(_y\)) orbitals of carbon atoms, exhibit a rigid red shift of ${\sim\,1}$ eV  and ${\sim\,0.5}$ eV  for \(100\%\) cases and  lower-concentration cases, respectively. The above-mentioned important results directly reflect the critical mechanisms, namely, the multi-orbital hybridizations of sp\(^3\)-p in Si-C bonds,  sp\(^3\)-sp\(^3\) in Si-Si bonds and sp\(^2\)-sp\(^2\) in C-C bonds. Such chemical bondings consist of four Si-orbitals and one C-orbital, being closely related to the bridge-site adsorption positions, and the binding energies $\&$ charge distributions of the separated orbitals. High-resolution ARPES measurements are very useful for verifying the low-energy valence bands crossing the Fermi level along K$\Gamma$ and M$\Gamma$ and the rigid $\sigma$ bands initiated from the $\Gamma$ point. On the other hand, all  substitution cases result in semiconducting behavior with a finite or vanishing band gap. The Dirac-cone structure presents a deviation from the $\Gamma$ point, a strong distortion, or even a full destruction. The number of valence and conduction bands remains the same after the chemical substitutions; furthermore, they are co-dominated by the Si-guest and C-host atoms. Apparently, the main features of the band structures in Si-substituted graphene systems arise from the sp\(^2\)-sp\(^2\) $\&$ p-p orbital hybridizations in Si-C bonds.
 
 \bigskip
 A lot of van Hove singularities in the atom- and orbital-decomposed density of states, which mainly come from the band-edge states, are created by the Si-guest-atom chemisorption and substitution. A pristine monolayer graphene only presents a V-shaped structure with a vanishing value at the Fermi level, the logarithmically symmetric peaks at ${-2.4}$ eV and 1.8 eV, and a shoulder structure at ${-3.0}$ eV, in which the former two and the last one, respectively, correspond to the $\pi$ and $\sigma$ bondings of carbon atoms. The strong $\pi$-bonding  evidences are totally destroyed by any chemisorption cases except for the very diluted Si-adatoms. They are replaced by a finite density of states at ${E_F=0}$ and many shoulder and peak structures. Furthermore, they are co-dominated by the four Si-(3s, 3p\(_x\), 3p\(_y\), 3p\(_z\)) orbitals and the single C-2p\(_z\) orbital, since their contributions are merged.  Specifically, the $\sigma$-band shoulder appears at ${-4.2}$ eV and ${-4.1}$ eV or ${-3.5}$ eV, and the ${1.0}$ eV or ${0.5}$ eV red shift is mainly determined by relative ionization energy of Si and C atoms. The above-mentioned significant features further support and illustrate the sp\(^3\)-p $\&$ sp\(^2\)-sp\(^2\) multi-orbital hybridizations in Si-C bonds and C-C bonds, respectively. As for the substitution cases, the density of states is zero within a finite energy range centered at the Fermi level, except for few systems with deformed V-shaped structures at $E_F$. All the Si-substituted graphene systems are finite- or zero-gap semiconductors. The special structures, which originate from Si-3p\(_z\) and C-2p\(_z\) orbitals, appear simultaneously. Also, a similar behavior is revealed in  Si-(3p\(_x\), 3p\(_y\)) and C-(2p\(_x\), 2p\(_y\)) orbitals. These clearly indicate the critical single- and multi-orbital hybridizations in Si-C bonds. The above-mentioned theoretical predictions could be verified by high-resolution STS measurements.
 
\bigskip
 \par\noindent {\bf Conflict of interest}
   
   There is no conflicts of interest in this paper.
          
\bigskip
  \par\noindent {\bf Acknowledgments}

This work was financially supported by the Ministry of Science and Technology (MOST) under grant number 108-2112-M-006-022-MY3, and the Hierarchical Green-Energy Materials (Hi-GEM) Research Center, from The Featured Areas Research Center Program within the framework of the Higher Education Sprout Project by the Ministry of Education (MOE) and the MOST (108-3017-F-006 -003) in Taiwan.

\newpage
\centering


\begin{thebibliography}{99}


            
    
   \bibitem{1} Fayos, J.  Possible 3D carbon structures as progressive intermediates in graphite to diamond phase transition.  \textit{Journal of Solid State Chemistry}, \textbf{148}, 278-285(1999).
                      
\bibitem{2} Wang, Y., Panzik, J. E., Kiefer, B. \(\&\) Lee, K. K.  Crystal structure of graphite under room-temperature compression and decompression. \textit{Scientific reports}, \textbf{2}, 520(2012).                    
                      
   \bibitem{3}  Novoselov, K. S.  et al.  Electric field effect in atomically thin carbon films. \textit{Science}, \textbf{306}, 666-669(2004).  
   
  \bibitem{4}  Zhang, Y., Tan, Y. W., Stormer, H. L. \(\&\) Kim, P.  Experimental observation of the quantum Hall effect and Berry's phase in graphene. \textit{Nature}, \textbf{438}, 201(2005).
            
   \bibitem{5}  Jiao, L., Wang, X., Diankov, G., Wang, H. \(\&\) Dai, H.  Facile synthesis of high-quality graphene nanoribbons. \textit{Nature nanotechnology}, \textbf{5}, 321(2010).     
   
   \bibitem{6}  Kosynkin, D. V. et al.  Longitudinal unzipping of carbon nanotubes to form graphene nanoribbons. \textit{Nature}, \textbf{458}, 872(2009). 
   
  \bibitem{7}  Cai, J. et al.  Atomically precise bottom-up fabrication of graphene nanoribbons. \textit{Nature}, \textbf{466}, 470(2010).  
  
\bibitem{8}  Issi, J. P., Langer, L., Heremans, J. \(\&\) Olk, C. H.  Electronic properties of carbon nanotubes: experimental results. \textit{Carbon}, \textbf{33}, 941-948(1995).

\bibitem{9}  Wilder, J. W., Venema, L. C., Rinzler, A. G., Smalley, R. E. \(\&\) Dekker, C.  Electronic structure of atomically resolved carbon nanotubes. \textit{Nature}, \textbf{391}, 59(1998).

\bibitem{10} Odom, T. W., Huang, J. L., Kim, P. \(\&\) Lieber, C. M.  Atomic structure and electronic properties of single-walled carbon nanotubes. \textit{Nature}, \textbf{391}, 62(1998).

 \bibitem{11} Yang, L., Jiang, J. \(\&\) Dong, J.  Formation mechanism of toroidal carbon nanotubes.  \textit{Physica status solidi (b)}, \textbf{238}, 115-119(2003).
 
\bibitem{12} Rocha, C. G., Pacheco, M., Barticevic, Z. \(\&\) Latg{\'e}, A.  Carbon nanotube tori under external fields. \textit{Physical Review B}, \textbf{70}, 233402(2004).
  
   \bibitem{13}  Howard, J. B., McKinnon, J. T., Makarovsky, Y., Lafleur, A. L., \(\&\) Johnson, M. E. Fullerenes C60 and C70 in flames. \textit{Nature}, \textbf{352}, 139(1991).  
    
 \bibitem{14}  Reinert, L. et al . Dispersion analysis of carbon nanotubes, carbon onions, and nanodiamonds for their application as reinforcement phase in nickel metal matrix composites. \textit{Rsc Advances}, \textbf{5}, 95149-95159.(2015).
 
\bibitem{15}  Park, H. J., Meyer, J., Roth, S. \(\&\) Sk{\'a}kalov{\'a}, V.  Growth and properties of few-layer graphene prepared by chemical vapor deposition. \textit{Carbon}, \textbf{48}, 1088-1094(2010).
 
\bibitem{16}  Kim, J., Lee, G. \(\&\) Kim, J.  Wafer-scale synthesis of multi-layer graphene by high-temperature carbon ion implantation. \textit{Applied Physics Letters}, \textbf{107}, 033104(2015).
 
\bibitem{17} Nagashio, K., Nishimura, T., Kita, K. \(\&\) Toriumi, A.  Mobility variations in mono-and multi-layer graphene films. \textit{Applied physics express}, \textbf{2}, 025003(2009).

\bibitem{18}  Latil, S. \(\&\) Henrard, L.  Charge carriers in few-layer graphene films. \textit{Physical Review Letters}, \textbf{97}, 036803(2006).
 
 \bibitem{19} Zhang, F. et al.  Spontaneous quantum Hall states in chirally stacked few-layer graphene systems. \textit{Physical review letters}, \textbf{106}, 156801(2011).     
 
\bibitem{20}  Ho, J. H., Lai, Y. H., Tsai, S. J., Hwang, J., Chang, C. \(\&\) Lin, M.F  Magnetoelectronic Properties of a Single-Layer Graphite (Condensed matter: electronic structure and electrical, magnetic, and optical properties).\textit{Journal of the Physical Society of Japan}, \textbf{75}, (2006).
        
 \bibitem{21}  Huang, Y. K., Chen, S. C., Ho, Y. H., Lin, C. Y. \(\&\) Lin, M. F.  Feature-rich magnetic quantization in sliding bilayer graphenes. \textit{Scientific reports}, \textbf{4}, 7509(2014). 
 
\bibitem{22} Koshino, M. \(\&\) McCann, E.  Landau level spectra and the quantum Hall effect of multilayer graphene. \textit{Physical Review B}, \textbf{83}, 165443(2011).

 \bibitem{23}  Wang, Z. F., Liu, F. \(\&\) Chou, M. Y.  Fractal Landau-level spectra in twisted bilayer graphene. \textit{Nano letters},  \textbf{12}, 3833-3838(2012).
 
 \bibitem{24}  Ho, J. H., Chang, C. P. \(\&\) Lin, M. F.  Electronic excitations of the multilayered graphite. \textit{Physics Letters A}, \textbf{352}, 446-450(2006).
 
\bibitem{25} Lin, M. F., Chuang, Y. C. \(\&\) Wu, J. Y.  Electrically tunable plasma excitations in AA-stacked multilayer graphene. \textit{Physical Review B},  \textbf{86}, 125434(2012).

\bibitem{26}  Wu, J. Y., Chen, S. C., Roslyak, O., Gumbs, G. \(\&\) Lin, M. F.  Plasma excitations in graphene: Their spectral intensity and temperature dependence in magnetic field. \textit{ACS nano}, \textbf{5}, 1026-1032(2011).

\bibitem{27} Lozovik, Y. E. \(\&\) Sokolik, A. A.  Influence of Landau level mixing on the properties of elementary excitations in graphene in strong magnetic field. \textit{Nanoscale research letters}, \textbf{7}, 134(2012).
 
 \bibitem{28} Ohta, T. et al.  Interlayer interaction and electronic screening in multilayer graphene investigated with angle-resolved photoemission spectroscopy. \textit{Physical Review Letters}, \textbf{98}, 206802(2007).
           
 \bibitem{29} Koshino, M. \(\&\) Ando, T.  Magneto-optical properties of multilayer graphene. \textit{Physical Review B}, \textbf{77}, 115313(2008).
  
  \bibitem{30}  Mucha-Kruczy{\'n}ski, M., Abergel, D. S. L., McCann, E. \(\&\) Fal{\'k}o, V. I.  On spectral properties of bilayer graphene: the effect of an SiC substrate and infrared magneto-spectroscopy. \textit{Journal of Physics: Condensed Matter}, \textbf{21}, 344206(2009).
    
  \bibitem{31}  Mucha-Kruczy{\'n}ski, M., McCann, E. \(\&\) Fal{\'k}o, V. I.  The influence of interlayer asymmetry on the magnetospectroscopy of bilayer graphene. \textit{Solid State Communications}, \textbf{149}, 1111-1116(2009).     
       
   \bibitem{32} Novoselov, K. S. et al. Two-dimensional gas of massless Dirac fermions in graphene. \textit{Nature}, \textbf{438}, 197(2005).
    
  \bibitem{33}  Chen, J. H., Jang, C., Xiao, S., Ishigami, M.  \(\&\) Fuhrer, M. S.  Intrinsic and extrinsic performance limits of graphene devices on SiO 2. \textit{Nature nanotechnology}, \textbf{3}, 206(2008).
       
   \bibitem{34}  Lee, C., Wei, X., Kysar, J. W. \(\&\)  Hone, J. Measurement of the elastic properties and intrinsic strength of monolayer graphene. \textit{Science}, \textbf{321}, 385-388(2008).  
   
 \bibitem{35}  Lee, J. K., et al.  The growth of AA graphite on (111) diamond. \textit{The Journal of chemical physics}, \textbf{129}, 234709(2008).
 
\bibitem{36} Jin, C., Lan, H., Peng, L., Suenaga, K. \(\&\) Iijima, S.  Deriving carbon atomic chains from graphene. \textit{Physical review letters}, \textbf{102}, 205501(2009).

\bibitem{37}  Lui, C. H.  et al.  Imaging stacking order in few-layer graphene. \textit{Nano letters}, \textbf{11}, 164-169(2010).

\bibitem{38}  Cancado, L. G.  et al.  Measuring the degree of stacking order in graphite by Raman spectroscopy. \textit{ Carbon}, \textbf{46}, 272-275(2008).

\bibitem{39}  Lui, C. H.  et al.  Observation of layer-breathing mode vibrations in few-layer graphene through combination Raman scattering.  \textit{Nano letters}, \textbf{12}, 5539-5544(2012).

\bibitem{40}  Zandiatashbar, A.  et al.   Effect of defects on the intrinsic strength and stiffness of graphene.  \textit{Nature communications}, \textbf{5}, 3186(2014). 

\bibitem{41}  Pereira, V. M. \(\&\) Neto, A. C.  Strain engineering of graphene’s electronic structure.  \textit{Physical Review Letters}, \textbf{103}, 046801(2009).

\bibitem{42}  Wong, J. H., Wu, B. R. \(\&\) Lin, M. F.  Strain effect on the electronic properties of single layer and bilayer graphene. \textit{The Journal of Physical Chemistry C}, \textbf{116}, 8271-8277(2012).


\bibitem{43}  Liu, Y., Liu, X., Zhang, Y., Xia, Q. \(\&\) He, J.  Effect of magnetic field on electronic transport in a bilayer graphene nanomesh. \textit{Nanotechnology}, \textbf{28}, 235303(2017).

\bibitem{44}  Lai, Y. H., Ho, J. H., Chang, C. P. \(\&\) Lin, M. F.  Magnetoelectronic properties of bilayer Bernal graphene. \textit{Physical Review B}, \textbf{77}, 085426(2008).

\bibitem{45}  Liu, H., Liu, Y. \(\&\) Zhu, D.  Chemical doping of graphene. \textit{Journal of materials chemistry}, \textbf{21}, 3335-3345(2011).

 \bibitem{46}  Chan, K. T., Neaton, J. B. \(\&\) Cohen, M. L.  First-principles study of metal adatom adsorption on graphene. \textit{Physical Review B}, \textbf{77}, 235430(2008).
 
 \bibitem{47} Garcia, J. C., de Lima, D. B., Assali, L. V. \(\&\) Justo, J. F.  Group IV graphene-and graphane-like nanosheets. \textit{ The Journal of Physical Chemistry C}, \textbf{115}, 13242-13246(2011).
 
 \bibitem{48} Liu, H. Y., Hou, Z. F., Hu, C. H., Yang, Y. \(\&\) Zhu, Z. Z.  Electronic and magnetic properties of fluorinated graphene with different coverage of fluorine. \textit{ The Journal of Physical Chemistry C}, \textbf{116}, 18193-18201(2012).
 
 
\bibitem{49}  Crook, C. B. et al.  Proximity-induced magnetism in transition-metal substituted graphene.  \textit{Scientific reports}, \textbf{5}, 12322(2015).

\bibitem{50}  Wu, J., Rodrigues, M. T. F., Vajtai, R. \(\&\) Ajayan, P. M.  Tuning the Electrochemical Reactivity of Boron-and Nitrogen-Substituted Graphene. \textit{Advanced Materials}, \textbf{28}, 6239-6246(2016).

\bibitem{51} Santos, E. J., Ayuela, A. \(\&\) S{\'a}nchez-Portal, D. First-principles study of substitutional metal impurities in graphene: structural, electronic and magnetic properties. \textit{New Journal of Physics}, \textbf{12}, 053012(2010).
 
\bibitem{52} Takahashi, T., Sugawara, K., Noguchi, E., Sato, T. \(\&\) Takahashi, T.  Band-gap tuning of monolayer graphene by oxygen adsorption. \textit{Carbon}, \textbf{73}, 141-145(2014).
 
  \bibitem{53} Sun, M.  et al.  First-principles study of the alkali earth metal atoms adsorption on graphene. \textit{Applied Surface Science}, \textbf{356}, 668-673(2015).
  
  \bibitem{54} Gao, H. \(\&\) Liu, Z.  DFT study of NO adsorption on pristine graphene. \textit{RSC Advances}, \textbf{7}, 13082-13091(2017).
  
\bibitem{55}  Yan, J. A. \(\&\) Chou, M. Y.  Oxidation functional groups on graphene: Structural and electronic properties. \textit{Physical review B}, \textbf{82}, 125403(2010). 
  
\bibitem{56} Cortes-Arriagada, D., Gutierrez-Oliva, S., Herrera, B., Soto, K. \(\&\) Toro-Labbe, A.  The mechanism of chemisorption of hydrogen atom on graphene: Insights from the reaction force and reaction electronic flux. \textit{The Journal of chemical physics}, \textbf{141}, 134701(2014).

\bibitem{57}  Lee, G., Lee, B., Kim, J. \(\&\) Cho, K.  Ozone adsorption on graphene: ab initio study and experimental validation. \textit{The Journal of Physical Chemistry C}, \textbf{113}, 14225-14229(2009).

\bibitem{58}  Ivanovskaya, V. V.  et al.  Hydrogen adsorption on graphene: a first principles study. \textit{The European Physical Journal B}, \textbf{76}, 481-486(2010).

\bibitem{59} Yan, H. J., Xu, B., Shi, S. Q.  \(\&\) Ouyang, C. Y.  First-principles study of the oxygen adsorption and dissociation on graphene and nitrogen doped graphene for Li-air batteries. \textit{Journal of Applied Physics}, \textbf{112}, 104316(2012).
   
\bibitem{60}  Beheshti, E., Nojeh, A. \(\&\) Servati, P.  A first-principles study of calcium-decorated, boron-doped graphene for high capacity hydrogen storage. \textit{Carbon}, \textbf{49}, 1561-1567(2011).
   
\bibitem{61}  Wu, M., Liu, E. Z., Ge, M. Y. \(\&\) Jiang, J. Z.  Stability, electronic, and magnetic behaviors of Cu adsorbed graphene: A first-principles study. \textit{Applied Physics Letters}, \textbf{94}, 102505(2009).


  \bibitem{62}  Lin, S. Y., Lin, Y. T., Tran, N. T. T., Su, W. P. \(\&\) Lin, M. F.  Feature-rich electronic properties of aluminum-adsorbed graphenes. \textit{Carbon}, \textbf{120}, 209-218(2017).


 \bibitem{63}  Tran, N. T. T., Nguyen, D. K., Glukhova, O. E. \(\&\) Lin, M. F.  Coverage-dependent essential properties of halogenated graphene: A DFT study. \textit{Scientific reports},  \textbf{7}, 17858(2017).
               
 \bibitem{64}  Hu, C. H.  et al.  Electronic and magnetic properties of silicon adsorption on graphene. \textit{Solid State Communications}, \textbf{151}, 1128-1130(2011).

\bibitem{65}  Yoshioka, T., Suzuura, H. \(\&\)  Ando, T.  Electronic states of BCN alloy nanotubes in a simple tight-binding model. \textit{Journal of the Physical Society of Japan}, \textbf{72}, 2656-2664(2003).

\bibitem{66}   da Rocha Martins, J. \(\&\) Chacham, H.  Disorder and segregation in B-C-N graphene-type layers and nanotubes: tuning the band gap. \textit{ACS Nano}, \textbf{5}, 385-393(2010).


\bibitem{67}  Cort{\'e}s-Arriagada, D., Miranda-Rojas, S., Ortega, D. E. \(\&\) Toro-Labb{\'e}, A.  Oxidized and Si-doped graphene: emerging adsorbents for removal of dioxane. \textit{Physical Chemistry Chemical Physics}, \textbf{19}, 17587-17597(2017).

\bibitem{68}  Ain, Q. T., Al-Modlej, A., Alshammari, A. \(\&\) Anjum, M. N.  Effect of solvents on optical band gap of silicon-doped graphene oxide.  \textit{Materials Research Express}, \textbf{5}, 035017(2018).

\bibitem{69}  Tang, Y. B. et al.  Tunable band gaps and p-type transport properties of boron-doped graphenes by controllable ion doping using reactive microwave plasma. \textit{ACS Nano}, \textbf{6}, 1970-1978(2012).
 
\bibitem{70}  Lin, S. Y., Chang, S. L., Tran, N. T. T., Yang, P. H. \(\&\) Lin, M. F.  H–Si bonding-induced unusual electronic properties of silicene: a method to identify hydrogen concentration. \textit{Physical Chemistry Chemical Physics}, \textbf{17}, 26443-26450(2015).   
      
\bibitem{71} Li, S. S.  et al.  Tunable electronic and magnetic properties in germanene by alkali, alkaline-earth, group III and 3d transition metal atom adsorption. \textit{Physical Chemistry Chemical Physics}, \textbf{16}, 15968-15978(2014).
  
 \bibitem{72}  Chen, R. B., Chen, S. C., Chiu, C. W. \(\&\) Lin, M. F.  Optical properties of monolayer tinene in electric fields. \textit{Scientific Reports}, \textbf{7},  1849(2017).
 
 
\bibitem{73}  Sahin, H. \(\&\) Peeters, F. M.  Adsorption of alkali, alkaline-earth, and 3d transition metal atoms on silicene.  \textit{Physical Review B}, \textbf{87}, 085423(2013).

\bibitem{74}  D{\'a}vila, M. E., Xian, L., Cahangirov, S., Rubio, A. \(\&\) Le Lay, G. Germanene: a novel two-dimensional germanium allotrope akin to graphene and silicene. \textit{New Journal of Physics}, \textbf{16}, 095002(2014).

\bibitem{75}  Cai, B.  et al.  Tinene: a two-dimensional Dirac material with a 72 meV band gap.  \textit{Physical Chemistry Chemical Physics}, \textbf{17}, 12634-12638(2015).


\bibitem{76}   Nie, S.  Scanning tunneling microscopy study of graphene on Au (111): Growth mechanisms and substrate interactions. \textit{Physical Review B}, \textbf{85},  205406(2012).


\bibitem{77}  Bir{\'o}, L.   Scanning tunnelling microscopy (STM) imaging of carbon nanotubes. \textit{Carbon}, \textbf{36},  689-696(1998).

\bibitem{78}  Chen, X., Wan, H., Song, K., Tang, D. \(\&\)  Zhou, G.  Scanning tunneling microscopy image modeling for zigzag-edge graphene nanoribbons. \textit{ Applied Physics Letters}, \textbf{98},  263103(2011).

\bibitem{79}  Wang, W. X.,   Scanning tunneling microscopy and spectroscopy of finite-size twisted bilayer graphene. \textit{Physical Review B}, \textbf{96},  115434(2017).

\bibitem{80}  Hattendorf, S., Georgi, A., Liebmann, M. \(\&\)  Morgenstern, M.  Networks of ABA and ABC stacked graphene on mica observed by scanning tunneling microscopy. \textit{Surface Science}, \textbf{610},  53-58(2013).

\bibitem{81}  Ortolani, L., Catheline, A., Morandi, V. \(\&\) P{\'e}nicaud, A.  Transmission Electron Microscopy Study of Graphene Solutions.  \textit{Springer, Berlin, Heidelberg}, \textbf{610},  157-163(2012).

\bibitem{82}  Kasumov, Y. A., Khodos, I. I., Kociak, M. \(\&\) Kasumov, A. Y.  Scanning and transmission electron microscope images of a suspended single-walled carbon nanotube. \textit{Applied physics letters}, \textbf{89}, 013120(2006).

\bibitem{83}  Elias, D. C. et al.  Control of graphene's properties by reversible hydrogenation: evidence for graphane. \textit{Science}, \textbf{323},  610-613(2009).

 \bibitem{84}  Kim, K. et al.   Multiply folded graphene.  \textit{Physical Review B}, \textbf{83},  245433(2011).

 \bibitem{85}  Xie, X. et al.   Controlled fabrication of high-quality carbon nanoscrolls from monolayer graphene. \textit{Nano letters}, \textbf{9},  2565-2570(2009).

\bibitem{86}   Liu, X. H. et al.  In situ transmission electron microscopy of electrochemical lithiation, delithiation and deformation of individual graphene nanoribbons. \textit{Carbon}, \textbf{50}, 3836-3844(2012).

 \bibitem{87} Warner, J. H., R{\"u}mmeli, M. H., Gemming, T., B{\"u}chner, B. \(\&\) Briggs, G. A. D.  Direct imaging of rotational stacking faults in few layer graphene.  \textit{Nano letters}, \textbf{9}, 102-106(2008).

\bibitem{88}   Norimatsu, W. \(\&\) Kusunoki, M.  Selective formation of ABC-stacked graphene layers on SiC (0001). \textit{Physical review B}, \textbf{81},  161410(2010).


 \bibitem{89} Luican, A., Li, G. \(\&\) Andrei, E. Y.  Quantized Landau level spectrum and its density dependence in graphene. \textit{Physical Review B}, \textbf{83}, 041405(2011).
 
\bibitem{90} Qiao, Z.  et al.  Quantum anomalous Hall effect in graphene from Rashba and exchange effects. \textit{Physical Review B}, \textbf{82}, 161414(2010).
 
\bibitem{91} Katsnelson, M. I.  Optical properties of graphene: The Fermi-liquid approach. \textit{EPL (Europhysics Letters)}, \textbf{84},  37001(2008).

\bibitem{92}  Kralj, M.  et al.  Graphene on Ir (111) characterized by angle-resolved photoemission.  \textit{Physical Review B}, \textbf{84}, 075427(2011). 

\bibitem{93}  Tran, N. T. T., Lin, S. Y., Lin, C. Y. \(\&\) Lin, M. F. Geometric and electronic properties of graphene-related systems: Chemical bonding schemes. \textit{CRC Press}, \textbf{book}, ISBN:9781138556522(2018).
               
\bibitem{94}   Lin, C. Y., Chen, R. B., Ho, Y. H. \(\&\) Lin, M. F.  Electronic and optical properties of graphite-related systems. \textit{CRC Press}, \textbf{book}, ISBN 9781138571068(2017).

\bibitem{95}  Son, Y. W., Cohen, M. L. \(\&\) Louie, S. G.  Energy gaps in graphene nanoribbons. \textit{Physical review letters}, \textbf{97}, 216803(2006).

\bibitem{96}  Ruffieux, P. et al.  Electronic structure of atomically precise graphene nanoribbons. \textit{ACS Nano}, \textbf{6}, 6930-6935(2012).

 
\bibitem{97} Senkovskiy, B. V. et al.  Finding the hidden valence band of N= 7 armchair graphene nanoribbons with angle-resolved photoemission spectroscopy. \textit{2D Materials}, \textbf{5}, 035007(2018).
        
\bibitem{98}  Nigar, S., Zhou, Z., Wang, H. \(\&\) Imtiaz, M.  Modulating the electronic and magnetic properties of graphene. \textit{RSC Advances}, \textbf{7}, 51546-51580(2017).    
        
\bibitem{99}   Medeiros, P. V., de Brito Mota, F., Mascarenhas, A. J. \(\&\) de Castilho, C. M.  Adsorption of monovalent metal atoms on graphene: a theoretical approach. \textit{Nanotechnology}, \textbf{21}, 115701(2010).    

\bibitem{100}  Lu, C. L., Chang, C. P., Huang, Y. C., Lu, J. M., Hwang, C. C. \(\&\) Lin, M. F.  Low-energy electronic properties of the AB-stacked few-layer graphites. \textit{Journal of Physics: Condensed Matter}, \textbf{18}, 5849(2006).
 

\bibitem{101} L. Lu, C., P. Chang, C., C. Huang, Y., H. Ho, J., C. Hwang, C. \(\&\) F. Lin, M.  Electronic properties of AA-and ABC-stacked few-layer graphites. \textit{Journal of the Physical Society of Japan}, \textbf{76}, 024701(2007).


\bibitem{102} Lin, M. F.  Low-frequency \(\pi\)-electronic excitations of simple hexagonal graphite. \textit{Journal of the Physical Society of Japan}, \textbf{70}, 897-901(2001).


\bibitem{103}  Choi, J., Lee, H. \(\&\) Kim, S.  Atomic-scale investigation of epitaxial graphene grown on 6H-SiC (0001) using scanning tunneling microscopy and spectroscopy. \textit{The Journal of Physical Chemistry C}, \textbf{114},  13344-13348(2010).

\bibitem{104}   Luican, A., Li, G. \(\&\) Andrei, E. Y.  Scanning tunneling microscopy and spectroscopy of graphene layers on graphite. \textit{Solid State Communications}, \textbf{149},  1151-1156(2009).

\bibitem{105}  LeRoy, B. J., Lemay, S. G., Kong, J. \(\&\) Dekker, C.  Scanning tunneling spectroscopy of suspended single-wall carbon nanotubes. \textit{Applied physics letters}, \textbf{84}, 4280-4282(2004).

\bibitem{106}  Odom, T. W., Huang, J. L., Kim, P., Ouyang, M. \(\&\) Lieber, C. M.  Scanning tunneling microscopy and spectroscopy studies of single wall carbon nanotubes. textit{Journal of materials research}, \textbf{13}, 2380-2388(1998).

\bibitem{107}  Niimi, Y.  et al.  Scanning tunneling microscopy and spectroscopy studies of graphite edges. \textit{Applied surface science}, \textbf{241},  43-48(2005).  


\bibitem{108} Sforzini, J., et al.  Structural and electronic properties of nitrogen-doped graphene. \textit{Physical review letters}, \textbf{116},  126805(2016).

\bibitem{109}  Dimakis, N.  et al.  Density functional theory calculations on alkali and the alkaline Ca atoms adsorbed on graphene monolayers. \textit{Applied surface science}, \textbf{413},  197-208(2017).

\bibitem{110}  Kong, X.  Metal-free Si-doped graphene: A new and enhanced anode material for Li ion battery. \textit{Journal of Alloys and Compounds}, \textbf{687}, 534-540(2016).

\bibitem{111}  Liu, X., Zhu, X. \(\&\) Pan, D.  Solutions for the problems of silicon-carbon anode materials for lithium-ion batteries. \textit{Royal Society Open Science}, \textbf{5},  172370(2018).

\bibitem{112}  Sivek, J., Sahin, H., Partoens, B. \(\&\) Peeters, F. M.  Adsorption and absorption of boron, nitrogen, aluminum, and phosphorus on silicene: Stability and electronic and phonon properties. \textit{Physical Review B}, \textbf{87},  085444(2013).

\bibitem{113}  Kaloni, T. P.  Tuning the structural, electronic, and magnetic properties of germanene by the adsorption of 3d transition metal atoms. \textit{The Journal of Physical Chemistry C}, \textbf{118}, 25200-25208(2014).

\bibitem{114}  Zhu, F. F.  et al.  Epitaxial growth of two-dimensional stanene. \textit{Nature materials}, \textbf{14},  1020(2015).

\bibitem{115}  Srivastava, P.  et al.   Tuning the electronic and magnetic properties of phosphorene by vacancies and adatoms. \textit{The Journal of Physical Chemistry C}, \textbf{119},  6530-6538(2015).

 \bibitem{116}   Akt{\"u}rk, E., Akt{\"u}rk, O. {\"U}. \(\&\) Ciraci, S. Single and bilayer bismuthene: Stability at high temperature and mechanical and electronic properties. \textit{Physical Review B}, \textbf{94},  014115(2016).
 
          
            \end{thebibliography}
\end{document}